%% file: Paper-3515.tex
\documentclass[runningheads]{llncs}
\usepackage{graphicx}
\usepackage{multirow}
\usepackage{amssymb}
\usepackage{wrapfig}
\usepackage{hyperref}
\usepackage{floatflt}
\usepackage{multirow,longtable}
\usepackage{diagbox} % Add the diagbox package
\usepackage{rotating}
\usepackage{xcolor}
\usepackage{amsmath}
\usepackage{amsfonts}

\begin{document}

\title{Estimation and Analysis of Slice Propagation Uncertainty in 3D Anatomy Segmentation}

\titlerunning{Slice Propagation Uncertainty}
\author{Rachaell Nihalaani \and Tushar Kataria \and Jadie Adams \and
Shireen Y. Elhabian \inst{3} }

% index{Nihalaani, Rachaell}
% index{Kataria, Tushar}
% index{Adams, Jadie}
% index{Y. Elhabian, Shireen}

\authorrunning{R Nihalaani et al.}
\institute{Kahlert School of Computing, University of Utah, Salt Lake City, USA \and
Scientific Computing and Imaging Institute, University of Utah, Salt Lake City, USA\and  Corresponding author\\ \{rachaellnihalaani,tushar.kataria,jadie,shireen\}@sci.utah.edu }

\maketitle              % typeset the header of the contribution
\input{abstract}
\input{introduction}

\input{methods-v2}
\input{results}
\input{conclusion}

%\clearpage

\bibliographystyle{splncs04.bst}

\bibliography{Paper-3515}
\newpage
\input{Supplementary}
\end{document}

%% file: abstract.tex
\begin{abstract}

Supervised methods for 3D anatomy segmentation demonstrate superior performance but are often limited by the availability of annotated data. This limitation has led to a growing interest in self-supervised approaches in tandem with the abundance of available unannotated data. 
Slice propagation has emerged as a self-supervised approach that leverages slice registration as a self-supervised task to achieve full anatomy segmentation with minimal supervision. This approach significantly reduces the need for domain expertise, time, and the cost associated with building fully annotated datasets required for training segmentation networks. 
However, this shift toward reduced supervision via deterministic networks raises concerns about the trustworthiness and reliability of predictions, especially when compared with more accurate supervised approaches.
To address this concern, we propose integrating calibrated uncertainty quantification (UQ) into slice propagation methods, which would provide insights into the model's predictive reliability and confidence levels.
Incorporating uncertainty measures enhances user confidence in self-supervised approaches, thereby improving their practical applicability.
We conducted experiments on three datasets for 3D abdominal segmentation using five UQ methods. 
The results illustrate that incorporating UQ improves not only model trustworthiness but also segmentation accuracy.
Furthermore, our analysis reveals various failure modes of slice propagation methods that might not be immediately apparent to end-users. This study opens up new research avenues to improve the accuracy and trustworthiness of slice propagation methods. 

\keywords{ Slice Propagation \and 3D Volume Segmentation \and Uncertainty Quantification \and Benchmarking \and Epistemic Uncertainty }

\end{abstract}

%% file: introduction.tex
\section{Introduction}

Achieving precise segmentation of 3D anatomy in MRI and CT volumes is critical for downstream tasks, including disease monitoring \cite{alnazer2021recent}, diagnostic processes \cite{icsin2016review}, and treatment planning \cite{kuisma2020validation}. 
Supervised deep learning models are state of the art for 3D segmentation of various anatomies and tumors \cite{niyas2022medical}. However, achieving high performance levels requires an enormous quantity of annotated image volumes \cite{li2023well}. 
Annotating these 3D volumes, which falls outside the scope of standard clinical routines, is both costly and time-intensive, as it necessitates the specialized knowledge of several radiologists, whose expertise is often in short supply.  
Therefore, it is crucial to develop techniques for 3D anatomy segmentation that can operate effectively with limited or no annotations.

Several machine-learning approaches have been developed for 3D anatomy segmentation with limited annotations.
\textit{Semi-supervised} methods (e.g., \cite{cai2023orthogonal,shi2021inconsistency}) use a combination of fully annotated volumes as well as un-annotated volumes to train deep learning models. These methods often utilize student-teacher frameworks, enhancing model performance through mechanisms such as consistency loss or mutual information maximization. 
Techniques such as pseudo-labeling and sample filtering are also used to augment the annotated data pool
to enhance the trained model \cite{arazo2020pseudo,swayamdipta2020dataset}. 
However, these methods still require high-quality, fully or partially annotated volumes for effective training and validation, limiting their utility in situations where annotated data are scarce or entirely unavailable.

\textit{Slice propagation} methods (e.g., \cite{bitarafan2022vol2flow,yeung2021sli2vol}) have evolved to offer a self-supervised approach for anatomy segmentation, eliminating the need for any annotated volumes for model training. 
Such methods use slice registration as a self-supervised task to establish correspondences between adjacent slices. 
During inference, only one annotated slice in the given volume is required to obtain a segmentation of the entire 3D anatomy. 
These methods reduce the burden on specialists of annotating entire volumes for training image segmentation networks. 
With a significant amount of available unannotated datasets, training networks on self-supervised slice-to-slice registration also significantly increases the training data, enhancing the model's ability to generalize. 
Moreover, these networks are trained to recognize pixel-wise correspondences of key low-level geometric features across various anatomies. As a result, they can adapt to a diverse range of anatomies, making them effective for inference on previously unseen anatomical structures. 
However, the performance of these self-supervised models is not yet comparable to that of supervised or semi-supervised methods trained on comparable quantities of data. Such a performance gap raises concerns about the accuracy, quality, and trustworthiness of model predictions. 
Given that deep learning models always produce outputs, irrespective of the confidence level of the prediction, it becomes crucial to quantify and analyze the reliability and accuracy of these self-supervised methods.
This scrutiny is particularly important in clinical scenarios where false negatives and false positives can critically impact patient outcomes. 

\textit{Uncertainty quantification (UQ)} is a key technique that helps identify when model predictions are reliable for clinical use or when extra caution is needed \cite{jungo2019assessing}. 
High predicted UQ values could flag potential incorrect segmented regions, guiding user interactions and refinement \cite{prassni2010uncertainty}. UQ analysis can also help identify failure modes of different methodologies that are not immediately apparent to the end-user. 
Uncertainty in deep learning predictions can be attributed to (a) the inherent uncertainty in the input data  (aleatoric uncertainty) and (b) the uncertainty in model parameters due to limited training data (epistemic uncertainty) \cite{kendall2017uncertainties}. 
Aleatoric uncertainty can be directly modeled as a function of the input data by producing probabilistic outputs. Epistemic uncertainty, on the other hand, is more difficult to quantify as it involves learning a distribution over model weights. However, quantifying epistemic uncertainty is crucial as it highlights the model's knowledge gaps and is directly correlated with the availability and diversity of the training data. Given that this paper focuses on weak/sparse supervision (through single-slice annotation), it is more relevant to quantify the limitations of the model's knowledge base using epistemic uncertainty quantification (UQ). This is clinically significant for conveying the trustworthiness of the results under limited supervision.
This paper makes the following contributions to the broader goal of establishing safe and trustworthy deep-learning models for medical applications:
\begin{itemize}
\vspace{-0.5em}
    \item The integration of epistemic UQ in self-supervised slice propagation methods, \textit{Sli2Vol} \cite{yeung2021sli2vol} and \textit{Vol2Flow} \cite{bitarafan2022vol2flow}, for analyzing the reliability and interpretability of anatomy segmentation.
    \item A comprehensive benchmark of five state-of-the-art methods for epistemic UQ on three datasets. \textit{Github Link}: \href{https://github.com/RachaellNihalaani/SlicePropUQ.git}{SlicePropUQ Github Repo.}
\end{itemize}

%% file: methods-v2.tex
\section{Uncertainty in Slice Propagation}

This section outlines the slice propagation techniques we consider in this paper and our adaptations to incorporate different UQ methods.
\vspace{-1em}
%%%%%%%%%%%%%%%%%%%%%%%%%%%%%%%%%%%%%%%%%%%%%%%%
\subsection{Slice Propagation Methods}
\textit{Sli2Vol} \cite{yeung2021sli2vol} and \textit{Vol2Flow} \cite{bitarafan2022vol2flow} are the state-of-the-art approaches for 3D volume segmentation using a single slice annotation. 
Let $\mathbf{I} \in \mathbb{R}^{H \times W \times D}$ denote a 3D image, where $H$, $W$ and $D$ are the height, width, and depth of the volume. 
These methods provide a segmentation of the entire volume propagating one manually annotated slice $\mathbf{S}_i \in \mathbb{R}^{H \times W}$, where $i \in \{1,\ldots, D\}$ to the entire 3D volume by learning robust pixel-wise correspondences between adjacent slices
in the volume via adjacency matrices.
For example, given two adjacent slices, $\mathbf{S}_i$ and $\mathbf{S}_{i+1}$, the models predict an affinity matrix $\mathbf{A}_{i+1, i}$. This matrix is then used to transform $\mathbf{S}_i$ to acquire an estimate of slice $(i+1)$, denoted $\hat{\mathbf{S}}_{i+1}$. The difference between the estimated slice, $\hat{\mathbf{S}}_{i+1}$, and the original slice, $\mathbf{S}_{i+1}$, is then used for model training. %Subsequent sections detail the differences between the methods. 
The following sections outline the method differences.

%%%%%%
\textbf{Sli2Vol \cite{yeung2021sli2vol}:}
In \textit{Sli2Vol} training, a pair of adjacent slices are first sampled from a training volume. An \textit{edge profile generator} is applied to the slices to extract edge features, followed by a convolutional neural network. An affinity matrix is then computed to capture the feature similarity between the two slices. The model is trained using a self-supervised mean square error (MSE) loss between the original slice and the slice reconstructed via the affinity matrix. During inference, the affinity matrices generated by the trained model are used to propagate the mask of the annotated slice throughout the volume iteratively. A verification module is also used to correct the propagated masks at each step, minimizing error accumulation and improving segmentation accuracy. 

% %%%%%%
\textbf{Vol2Flow \cite{bitarafan2022vol2flow}:}
Whereas \textit{Sli2Vol} utilizes a 2D CNN for slice-by-slice transformation, \textit{Vol2Flow} employs a 3D registration network to apply a sequence of transformations across the entire volume, enabling dense segmentation mask generation for each test volume. Vol2Flow's architecture is inspired by 3D-UNet networks, producing two sets of displacement deformation fields (DDFs) for forward and backward information propagation between adjacent slices. The learning process entails generating neighboring slices around each source slice using the DDFs and minimizing a boundary-preserving loss function that combines a structural similarity index and an edge-preserving loss. For mask propagation, Vol2Flow applies sequential transformations to generate pseudo labels for slices, introducing a refinement function to correct errors. The refinement method employs a non-linear classifier, specifically an SVM with an RBF kernel, to improve the classification of pixels during mask propagation.%, addressing the limitations of linear classification and reducing false positives.

%%%%%%%%%%%%%%%%%%%%%%%%%%%%%%%%%%%%%%%%%%%%%%%%
\subsection{Epistemic Uncertainty Quantification}
Quantification of epistemic uncertainty in deep learning models is challenging since it entails learning a distribution over model weights \cite{kendall2017uncertainties}. 
The variance in predictions made with weights sampled from such a distribution is directly proportional to the degree of model uncertainty, where low prediction variance signifies low uncertainty or high model confidence. 
Although scalable epistemic UQ techniques have been proposed and successfully applied to supervised segmentation models \cite{adams2023benchmarking}, extending these advancements to slice propagation techniques has not been studied. 
The next section describes five state-of-the-art epistemic UQ methods and their adaptation for slice propagation tasks.

%%%%%%
\textbf{Deep Ensemble \cite{lakshminarayanan2017simple}}
enhances prediction accuracy and robustness by leveraging multiple independently initialized, identically trained models (e.g., ensemble members) \cite{lakshminarayanan2017simple}.
In this frequentist approach to UQ, the predictions from each ensemble member provide a distribution.
The mean of this distribution provides a robust ensemble prediction, and the variance captures epistemic uncertainty. We used four initializations to train slice propagation models, namely base initialization (similar to original implementations), Kaiming (He) uniform initialization, Glorot (Xavier) uniform initialization, and a custom normal initialization. This strategy ensures each model in the ensemble explores different data representations, encouraging prediction diversity for calibrated UQ and enhancing the ensemble's generalization capabilities. 

%%%%%%
\textbf{Batch Ensemble \cite{wen2020batchensemble}} improves on the traditional deep ensemble approach by significantly reducing the computational and memory demands, which typically scale linearly with the number of ensemble members. 
Batch ensemble compromises between a single network and a full ensemble by defining member-specific convolutional weight matrix $\mathbf{W}_i$, as the Hadamard product of a shared base weight matrix $\mathbf{W}_{\text{shared}}$ and two member-specific rank-one vectors, $\mathbf{r}_i$ and $\mathbf{s}_i$: $\mathbf{W}_i = \mathbf{r}_i \circ \mathbf{W}_{\text{shared}} \circ \mathbf{s}_i^T$.  
When training slice propagation networks with the batch ensemble, predictions are made with each set of member-specific weights. Then, similar to deep ensemble, member predictions are averaged in inference to provide a robust prediction, and the variance in predictions provides UQ. 

%%%%%
\textbf{Monte Carlo (MC) Dropout \cite{gal2016dropout}} is a widely used regularization technique that also provides a scalable solution to approximate variational inference \cite{gal2016dropout}. 
MC Dropout involves randomly omitting model weights during training and inference, enabling the model to produce a range of predictions for the same input. This variability captures the degree of model certainty, as confident models will make similar predictions with different dropout masks. For slice propagation methods, we place a dropout layer within each network block after ReLU activation \cite{adams2020uncertain}. The dropout rate was selected to be 0.2 in hyperparameter tuning to balance accuracy and model robustness.

%%%%%%
\textbf{Concrete Dropout \cite{gal2017concrete}}. MC dropout requires time-consuming and computationally expensive manual tuning of layer-wise dropout rates to acquire well-calibrated UQ. However, concrete dropout employs a continuous relaxation of the dropout's discrete masks, allowing for the automatic optimization of per-layer dropout probabilities in tandem with network weights, significantly streamlining the process. We integrated spatial concrete dropout within all convolutional layers of slice propagation methods.

%%%%%%
\textbf{Stochastic Weight Averaging Gaussian (SWAG) \cite{maddox2019simple}}
builds upon stochastic weight averaging \cite{izmailov2018averaging}, a technique that defines model weights as the average of weights traversed during stochastic gradient descent (SGD) after initial convergence to find a broader optimum. SWAG fits a Gaussian distribution over these traversed weights to model the posterior distribution of network weights. The posterior estimation facilitates the generation of a distribution of predictions that capture model uncertainty. The mean weights and their covariance matrices are obtained during post-convergence training to estimate the Gaussian weight distribution for slice propagation. We then sampled this distribution to obtain the varied predictions needed for uncertainty estimation in inference.

The selection of 5 diverse SOTA scalable epistemic UQ methods is informed by their varied approaches (covering frequentist and Bayesian perspectives) to addressing model uncertainty effectively \cite{adams2023benchmarking}. Each method was chosen for its unique strengths: Deep and Batch Ensembles provide robustness through model averaging; MC Dropout and Concrete Dropout facilitate practical uncertainty estimation during training and inference, critical for deployment in clinical environments; and SWAG captures variability in model parameters through its approximation of the posterior distribution. This diverse toolkit allows us to comprehensively evaluate and enhance the predictive reliability and interpretability of self-supervised slice propagation methods. 

%\vspace{0.5em}
%%%%%%%%%%%%%%%%%%%%%%%%%%%%%%%%%%%%%%%%%%%%%%%%
\noindent {\textbf{Evaluation Metrics}.} To evaluate the performance of slice propagation methods, we utilize the \textbf{Dice Similarity Coefficient (DSC)}\cite{kataria2023automating,chen2017rethinking}. Additionally, to check if the predicted segmentation conforms to the actual organ boundaries and surfaces, we also report \textbf{Surface Dice} and \textbf{Average Hausdorff Distance (AHD)}. 
To assess the UQ calibration, we consider the \textbf{Pearson correlation coefficient} ($r$) between the predicted epistemic uncertainty and prediction error (1-DSC). 
A higher $r$ value signifies better UQ calibration, as we would expect model uncertainty to be high when the prediction error is high. 
 We also utilize the \textbf{area under the error retention curve (R-AUC)} to jointly assess segmentation accuracy and UQ calibration.
Error-retention curves plot the prediction error (100 - DSC) against the proportion of data retained after iteratively excluding predictions with the highest uncertainty. The R-AUC quantifies the model's performance across different levels of uncertainty retention. A lower R-AUC indicates better accuracy since it implies lower error across retained predictions, and better UQ calibration since it implies uncertainty/error correlation.

%% file: results.tex
\section{Results}

\textbf{Datasets Used.} 
Our study employs a diverse array of datasets for training and evaluation. 
We sourced our training data from: KiTS \cite{heller2023kits21}, which includes 300 multi-phase CT scan volumes; CT Lymph Nodes \cite{ct-lymphnodes}, utilizing the 86 abdominal lymph node volumes; and Pancreas-CT \cite{ct-pancreas}, with 82 volumes of abdominal contrast-enhanced 3D CT scans.
For evaluation, we used three datasets: SLIVER07 \cite{heimann2009comparison}, consisting of 20 clinically sourced 3D CT liver volumes; CHAOS \cite{valindria2018multi}, a comprehensive multi-organ segmentation dataset where we used only CT-based liver segmentation; and DecathSpleen \cite{simpson2019large}, featuring 41 volumes of portal venous phase 3D CT scans for spleen segmentation.

%%%%%%%%%%%%%%%%%%%%%%%%%%%%%%%%%%%%%%%%%%%%%%%%
\noindent \textbf{Implementation Details.} 
We used the original implementations of Sli2Vol \cite{yeung2021sli2vol} and Vol2Flow \cite{bitarafan2022vol2flow}.  
For UQ methods on 3D volume segmentation, we referenced the implementation release by a recent UQ benchmark \cite{adams2023benchmarking}.
We scale the slice featuring the anatomy's largest manually annotated area as the annotated slice.
We used four ensemble members for ensemble-based models and 30 posterior samples to get an average prediction for dropout and SWAG methods. We use only four ensemble members based on empirical findings, balancing computational (limited GPU memory) requirements and performance gain.
\vspace{-1em}
\subsection{Uncertainty Quantification Results}\label{ssec:results:sli2vol}

\textbf{UQ Methods Performance Insights.} 
Table \ref{tab:sli2vol} presents comprehensive results of UQ methods applied across three datasets for both Sli2Vol and Vol2Flow. Most UQ methods improve upon baseline performances in DSC and surface metrics, underscoring their utility in boosting model precision and reliability across varied datasets and tasks. 

Baseline deterministic models, without the integration of UQ, establish a fundamental level of accuracy, yet exhibit notable boundary detection challenges, as indicated by higher AHD values. Concrete dropout demonstrates superior segmentation and uncertainty estimation, outperforming other UQ methods in accuracy. %
SWAG shows higher UQ calibration than ensemble methods, even with low-accuracy models, indicating that SWAG is a good UQ estimator, which will be important for annotation-efficient models. The SWAG method results in reduced segmentation accuracy, likely because it utilizes an average of converged weights, whereas the baseline employs the best of converged weights as assessed by the validation performance. However, SWAG provides better calibrated UQ than other methods, which suggests that it learns a wide posterior distribution of weights, resulting in diverse predictions that provide a poorer averaged estimate but accurately capture the degree of model confidence.

The results highlight the efficacy of dropout techniques, specifically concrete dropout, for their robust segmentation and uncertainty quantification capabilities. Combined with the ease of integration, these robust capabilities make dropout highly effective for medical imaging tasks. Deep ensemble and batch ensemble offer a balanced improvement over baselines, and SWAG's notable uncertainty estimation ability underscores the need for careful UQ method selection based on specific application requirements. This analysis emphasizes the critical role of choosing the right UQ approach to optimize model performance, considering the trade-offs between reducing errors and enhancing correlation measures.

\begin{table}
\vspace{-1em}
\caption{\textbf{UQ results}. Mean and standard deviations of DSC, surface dice, AHD, r, and R-AUC scores using predictions from Sli2Vol \cite{yeung2021sli2vol} and Vol2Flow \cite{bitarafan2022vol2flow} on all three datasets.
}
\centering
\small
\setlength{\tabcolsep}{2pt} % Adjust the spacing between columns
\scalebox{0.56}{
\begin{tabular}{p{0.25cm}|c|ccccc||ccccc}
& & \multicolumn{5}{c||}{\bf Sli2Vol} &  \multicolumn{5}{c}{\bf Vol2Flow} \\
\hline
%& & & & & Average & & & & Average   \\
& UQ Methods    & DSC $\uparrow$ & Surface Dice $\uparrow$ & AHD $\downarrow$ & $r$ $\uparrow$ & R-AUC $\downarrow$ 
                & DSC $\uparrow$ & Surface Dice $\uparrow$ & AHD $\downarrow$ & $r$ $\uparrow$ & R-AUC $\downarrow$  \\
\hline
\multirow{6}{=}{\begin{sideways}\bf SLiver07 \end{sideways}}
& Base (w/o UQ)             & 91.44$\pm$2.94    & 65.92$\pm$6.59    & 110.70$\pm$54.92  & -     & -
                    & 92.58$\pm$3.68    & 69.95$\pm$4.32    & 95.34$\pm$35.24  & -     & -\\
& Deep Ensemble     & 91.68$\pm$3.02    & 67.53$\pm$7.52    & 99.32$\pm$47.36   & 0.13  & 4.09$\pm$1.38
                    & 92.97$\pm$4.15    & 71.54$\pm$5.21    & 89.47$\pm$32.88   & 0.27  & 3.25$\pm$2.47     \\
& Batch Ensemble    & 91.08$\pm$2.91    & 65.67$\pm$8.27    & 109.32$\pm$54.14   & 0.08  & 4.48$\pm$1.35
                    & 92.13$\pm$3.77    & 70.06$\pm$5.98    & 95.15$\pm$38.75   & 0.15  & 2.95$\pm$3.21     \\
& MC Dropout        & 92.40$\pm$2.89    & 67.52$\pm$9.28    & 101.51$\pm$43.51  & 0.19  & 3.30$\pm$1.12
                  & 93.58$\pm$3.41    & 71.45$\pm$6.89    & 90.84$\pm$27.63  & 0.35  & 3.22$\pm$3.52     \\
& Concrete Dropout  & \textbf{92.48$\pm$3.16}    & \textbf{68.82$\pm$9.45}    & 93.27$\pm$43.84   & 0.23  & 3.49$\pm$1.26
                   & \textbf{94.02$\pm$1.98}    & \textbf{72.49$\pm$5.66}    & \textbf{85.48$\pm$29.55}   & 0.43  & \textbf{2.94$\pm$2.81}     \\
& SWAG              & 83.03$\pm$7.33    & 39.97$\pm$1.39    & \textbf{48.62$\pm$23.15 }  & \textbf{0.69}  & \textbf{3.21$\pm$1.59}
                    & 85.26$\pm$5.74    & 61.52$\pm$9.65    & 93.84$\pm$17.81   & \textbf{0.67}  & 3.17$\pm$3.88     \\
\hline
\multirow{6}{=}{\begin{sideways} \bf DecathSpleen\end{sideways}} 
& Base (w/o UQ)             & 89.41$\pm$8.77            & 85.97$\pm$10.54           & 19.39$\pm$11.84           &-              &-                  
                   & 86.55$\pm$7.29            & 82.24$\pm$8.67            & 15.56$\pm$8.68            &-              &-                          \\
& Deep Ensemble     & 89.97$\pm$8.49            & 87.42$\pm$10.51           & 16.43$\pm$12.01           & 0.15          & 5.88$\pm$5.44     
                    & 87.15$\pm$7.06            & 84.98$\pm$7.45            & 14.74$\pm$9.25            & 0.22          & 5.12$\pm$3.97             \\
& Batch Ensemble    & 88.97$\pm$8.73            & 86.06$\pm$11.04           & 19.16$\pm$15.85           & 0.38          & 6.36$\pm$5.42     
                   & 86.22$\pm$7.87            & 84.00$\pm$8.56            & 15.02$\pm$10.57           & 0.40          & 5.99$\pm$4.52             \\
& MC Dropout        & 90.78$\pm$6.85            & 87.89$\pm$9.80            & 14.67$\pm$10.40           & \textbf{0.52} & 4.84$\pm$4.57     
                   & 89.24$\pm$8.24            & 85.45$\pm$10.57           & 13.95$\pm$8.65            & \textbf{0.58} & 4.78$\pm$5.48             \\
& Concrete Dropout  & \textbf{91.24$\pm$6.83}   & \textbf{88.80$\pm$9.70}   & \textbf{13.78$\pm$9.44}   & 0.27          & 5.05 $\pm$ 4.88   
                    & \textbf{90.82$\pm$6.54}   & \textbf{87.65$\pm$9.21}   & \textbf{12.84$\pm$9.87}   & 0.32          & \textbf{4.56 $\pm$ 5.35}  \\
& SWAG              & 84.55$\pm$9.12            & 73.00$\pm$16.75           & 15.82$\pm$6.56            & 0.42          & \textbf{4.28 $\pm$ 2.69}   
                    & 81.06$\pm$10.41           & 74.58$\pm$11.56           & 13.24$\pm$10.45           & 0.41          & 4.92 $\pm$ 3.58           \\      
\hline
\multirow{6}{=}{\begin{sideways}\bf CHAOS\end{sideways}} 
& Base (w/o UQ)             & 91.11$\pm$8.82    & 41.96$\pm$13.86   & 67.49$\pm$21.41   &- &-  
                    & 85.31$\pm$4.62    & 46.54$\pm$9.21    & 62.84$\pm$15.55   &- &-  \\
& Deep Ensemble     & \textbf{93.66$\pm$7.21}    & \textbf{50.70$\pm$12.96}   & 65.64$\pm$22.40   & 0.75  & 7.14$\pm$5.23     
                    & 86.17$\pm$4.35    & 48.62$\pm$8.91    & 59.98$\pm$17.24   & 0.77  & 6.53$\pm$6.11     \\
& Batch Ensemble    & 92.30$\pm$6.59    & 43.20$\pm$12.43    & 67.24$\pm$24.25   & 0.80  & 7.40$\pm$4.64
                    & 86.57$\pm$5.89    & 47.56$\pm$9.11    & 62.08$\pm$20.57   & 0.82  & 6.59$\pm$5.22     \\
                    
& MC Dropout        & 93.40$\pm$6.77    & 45.29$\pm$12.77   & 60.14$\pm$20.93   & 0.75  & 6.53$\pm$4.60     
                    & \textbf{87.09$\pm$5.61}    & 51.65$\pm$8.84    & 57.84$\pm$14.65   & 0.81  & \textbf{5.89$\pm$3.84}     \\
& Concrete Dropout  & 92.36$\pm$8.02    & 45.37$\pm$13.11   & 57.67$\pm$20.38   & \textbf{0.81}  & 7.68$\pm$5.90
                   & 86.82$\pm$4.89    & \textbf{54.89$\pm$10.65}   & \textbf{54.58$\pm$13.64}   & \textbf{0.87}  & 6.12$\pm$4.55     \\
& SWAG              & 84.32$\pm$6.79    & 32.70$\pm$7.29    & \textbf{48.32$\pm$10.62}   & 0.42  & \textbf{4.20$\pm$1.63}
                    & 74.41$\pm$5.22    & 42.26$\pm$8.47    & 58.41$\pm$18.54   & 0.51  & 5.94$\pm$4.87     \\    
\hline
\end{tabular}}
\label{tab:sli2vol}
\vspace{-2em}
\end{table}

\begin{figure}[!htb]
    \centering
    \includegraphics[scale=0.14]{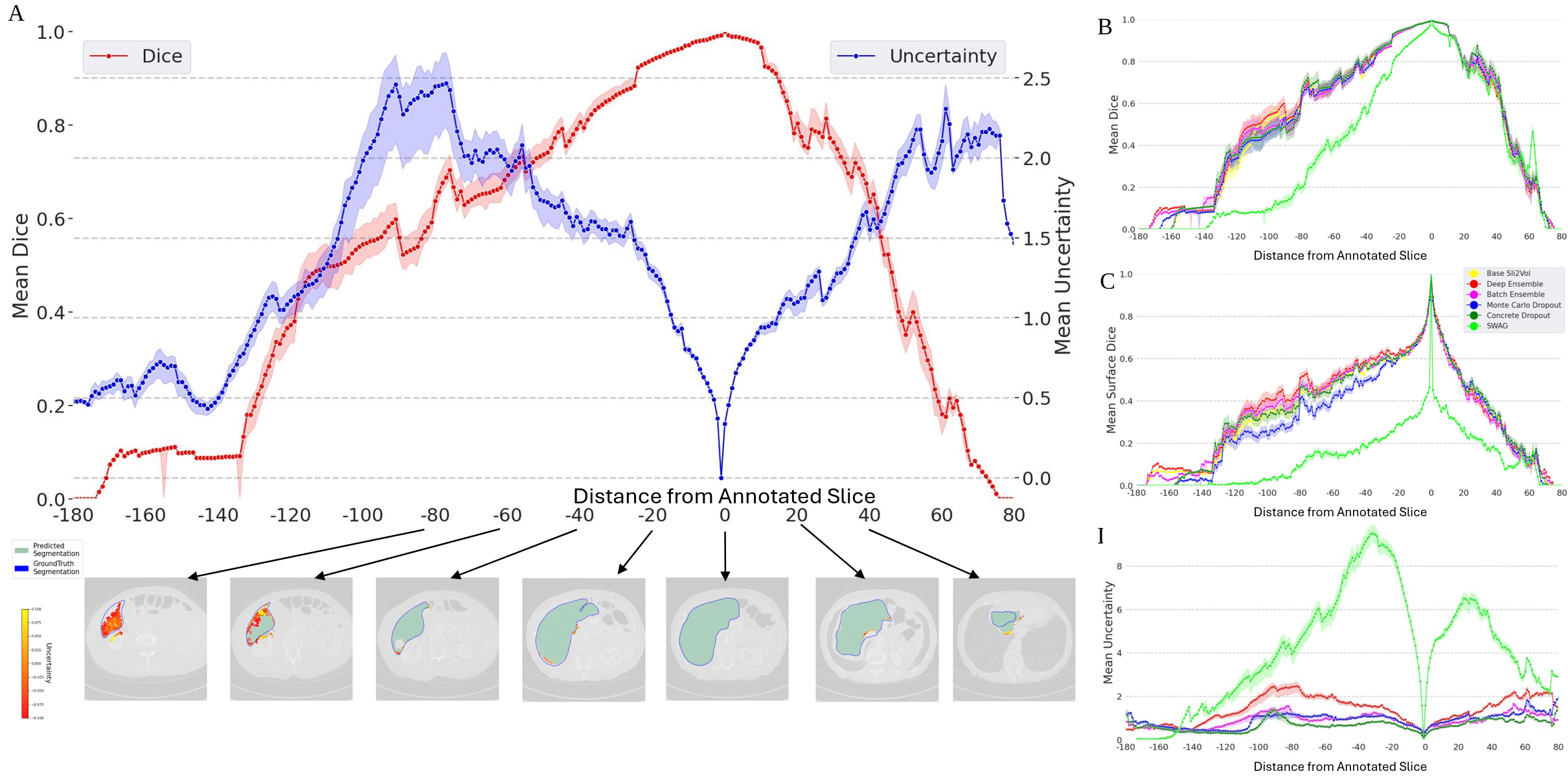}
    \caption{\textbf{Sli2Vol accuracy and uncertainty variation as a function of distance from annotated slice} A. Comparative analysis of variability in DSC and uncertainty metrics relative to the distance from the annotated slice when using \textbf{concrete dropout} (dataset: DecathSpleen). Performance metrics (B) \textbf{DSC},(C) \textbf{surface dice} and (D) \textbf{uncertainty} for all UQ Methods, relative to the distance from the annotated slice.}
    \label{fig:sli2vol:uncertainity-epoch}
    \vspace{-1em}
\end{figure}

\noindent\textbf{Performance Trends.} 
In assessing the Sli2Vol and Vol2Flow models, we also report metrics on a more granular slice level, shedding light on critical knowledge gaps that are not captured by the mean performance statistics over the dataset.
Mean DSC and associated uncertainties, when averaged over large volumes, could mask the true performance nuances of these models. 
Our targeted analysis revealed a discernible performance drop and increased uncertainty in models as segmentation predictions deviate from the manually annotated slice, as shown in Figure \ref{fig:sli2vol:uncertainity-epoch}A. This trend reflects a model's diminishing accuracy and confidence as it ventures further from the annotated slice.
We observe that the error and uncertainty estimates correctly correlate across distances from the annotated slice, suggesting accurate confidence estimation. Figures \ref{fig:sli2vol:uncertainity-epoch}B and C bring to attention the steeper decline in surface dice scores as compared to overall DSC, signaling a quicker degradation in the model's ability to retain surface information shortly beyond the annotated slice. This quick degradation is problematic as it indicates a rapid loss of accuracy in capturing the intricate contours and edges of anatomical structures. In medical applications, such as surgical planning or tumor resection, the precise delineation of surfaces is crucial. Imprecise surface segmentation can lead to inaccurate identification of tumor margins or critical anatomical landmarks, compromising treatment outcomes. A similar pattern emerged in the Vol2Flow analysis, shown in Figure \ref{fig:vol2flow:uncertainity-epoch}, affirming this as a broader trend across both modeling approaches. Supplementary GIFs are provided to visually demonstrate the progression of predicted segmentations and associated uncertainties through a volume.

\noindent\textbf{Failure Modes in Slice Propagation Methods.} Through detailed analysis of Figures \ref{fig:sli2vol:uncertainity-epoch} and \ref{fig:vol2flow:uncertainity-epoch}, along with observations from our study, we identify several key limitations of Sli2Vol and Vol2Flow models: (a) A pronounced decline in performance metrics is observed as soon as models predict slices merely 5-20 mm away from the annotated slice. Notably, the accuracy of surface metrics diminishes starting from slices closely adjacent to the annotated slice.
%almost immediately beyond the annotated slice. 
(b) The models exhibit difficulties in handling non-convex anatomical structures where the segmentation is discontinuous or due to branching anatomical structures. For example, across slices, femur bone segmentation becomes split into two substructures, the greater trochanter and the femoral head.   (c) Training via a surrogate registration task can inadvertently bias models toward assuming structural continuity since it emphasizes aligning continuous structures across slices. This bias results in the failure of the models to recognize the natural discontinuity or endpoints of anatomical features, resulting in over-extended segmentations. 

%% file: conclusion.tex
\section{Conclusion and Future Work} 
This study focuses on using slice propagation to reduce manual annotations in training and limit inference annotations to a single slice, addressing the trustworthiness of these methods amid minimal expert-driven supervision. We integrated five UQ methods to evaluate their accuracy and uncertainty calibration in medical image analysis. While we do not propose a novel method of epistemic UQ estimation, we provide open-source, non-trivial extensions of existing methods to this new task and benchmark their performance.
Our assessments demonstrate that incorporating UQ into slice propagation approaches enhances predictive accuracy and provides usable confidence estimation, effectively bridging the gap between semi-automatic methods and user reliance, previously unexplored. 
Furthermore, our investigation uncovers critical failure modes in slice propagation methods that may go unnoticed by users. Recognizing these shortcomings is pivotal to inform improvement and continuous model refinement.
Future work could explore the influence of domain variation (e.g., CT vs MRI) on uncertainty estimation.
Additionally, there is substantial potential to adapt UQ techniques to more adeptly handle the intricate challenges of slice propagation. 
Calibrated UQ could guide the development of methodologies capable of mitigating the failure modes observed in current segmentation methods.
This work reveals the potential and shortcomings of slice propagation segmentation models with UQ, increasing the potential for safe, feasible self-supervised anatomy segmentation. 

\section*{Acknowledgements} 
This work was supported by the National Institutes of Health under grant numbers NIBIB-U24EB029011 and NIAMS-R01AR076120. The content is solely the responsibility of the authors and does not necessarily represent the official views of the National Institutes of Health.

\subsubsection{\discintname} The authors have no competing interests to declare that are
relevant to the content of this article.
% We thank the Scientific Computing and Imaging Institute, and the Kahlert School of Computing at the University of Utah for their support of this Project. 
%

%% file: Supplementary.tex
%\section{Supplementary}

\begin{figure}[!htb]
    \centering
    \includegraphics[scale=0.14]{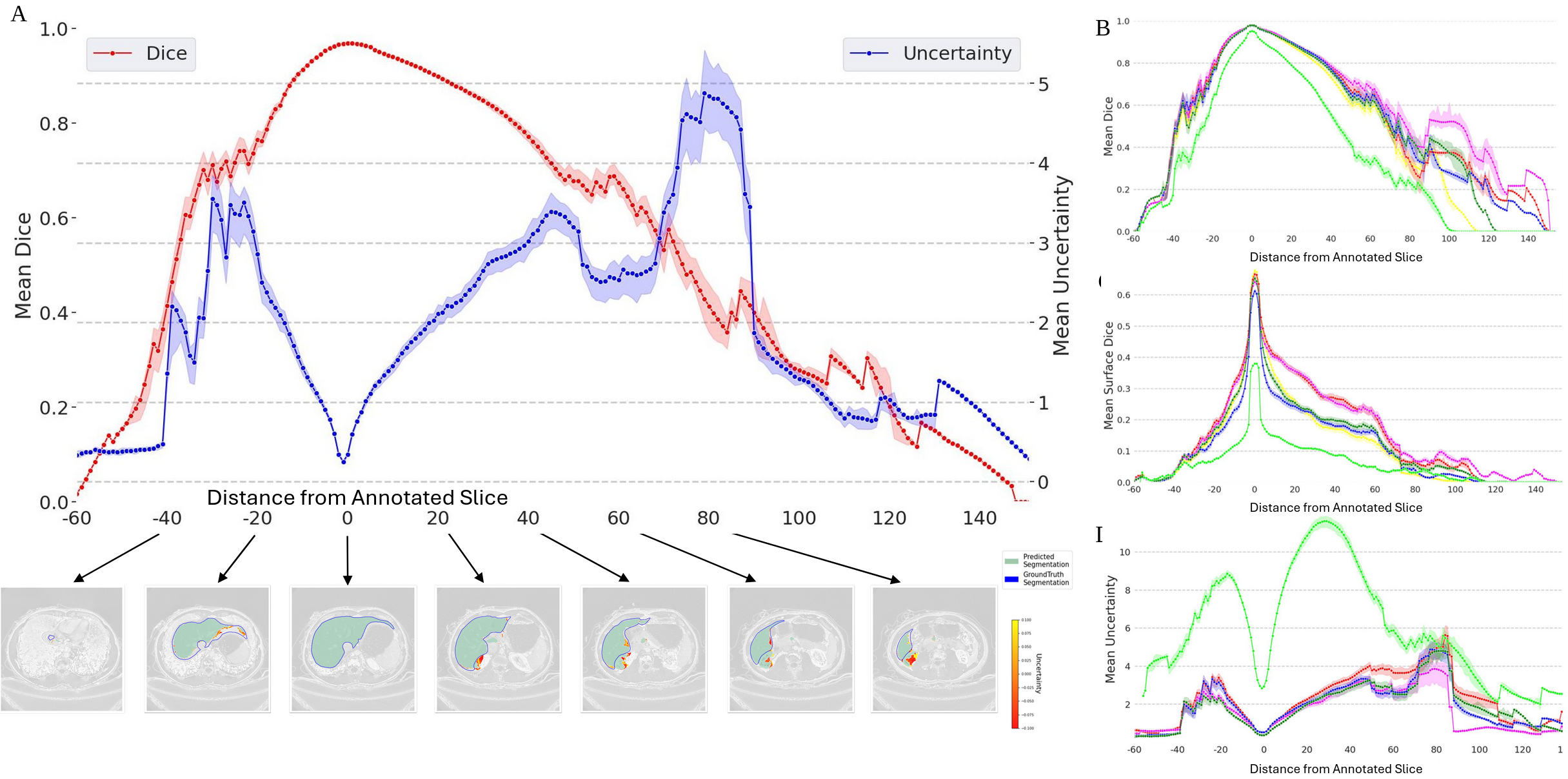}
    \caption{\textbf{Vol2Flow accuracy and uncertainty variation as function of distance from annotated slice} A. Comparative analysis of variability in DSC and uncertainty metrics relative to the distance from the annotated slice when using \textbf{MC dropout} (dataset: CHAOS). Performance metrics (B) \textbf{DSC},(C) \textbf{surface dice} and (D) \textbf{uncertainty} for all UQ Methods, relative to the distance from the annotated slice.}
    \label{fig:vol2flow:uncertainity-epoch}
    \vspace{-1.5em}
\end{figure}

\begin{figure}[!htb]
    \centering
    \includegraphics[scale=0.14]{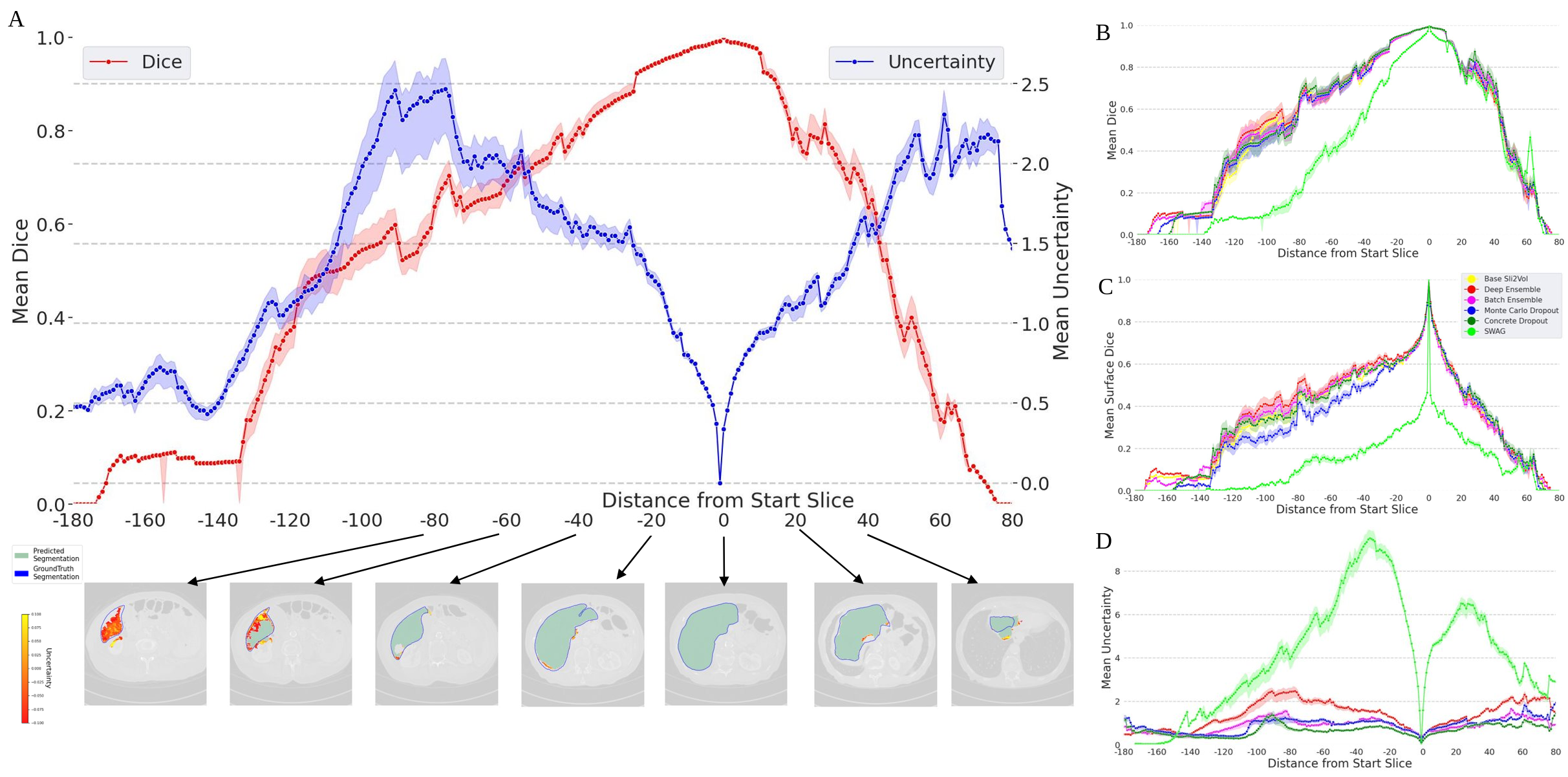}
    \caption{\textbf{Sli2Vol accuracy and uncertainty variation as function of distance from annotated slice} A. Comparative analysis of variability in DSC and uncertainty metrics relative to the distance from the annotated slice when using \textbf{concrete dropout} (dataset: SLiver07). Performance metrics (B) \textbf{DSC},(C) \textbf{surface dice} and (D) \textbf{uncertainty} for all UQ Methods, relative to the distance from the annotated slice.}
    \label{fig:sli2vol:uncertainity-epoch-2}
    \vspace{-1em}
\end{figure}

\begin{figure}[!htb]
    \centering
    \includegraphics[scale=0.16]{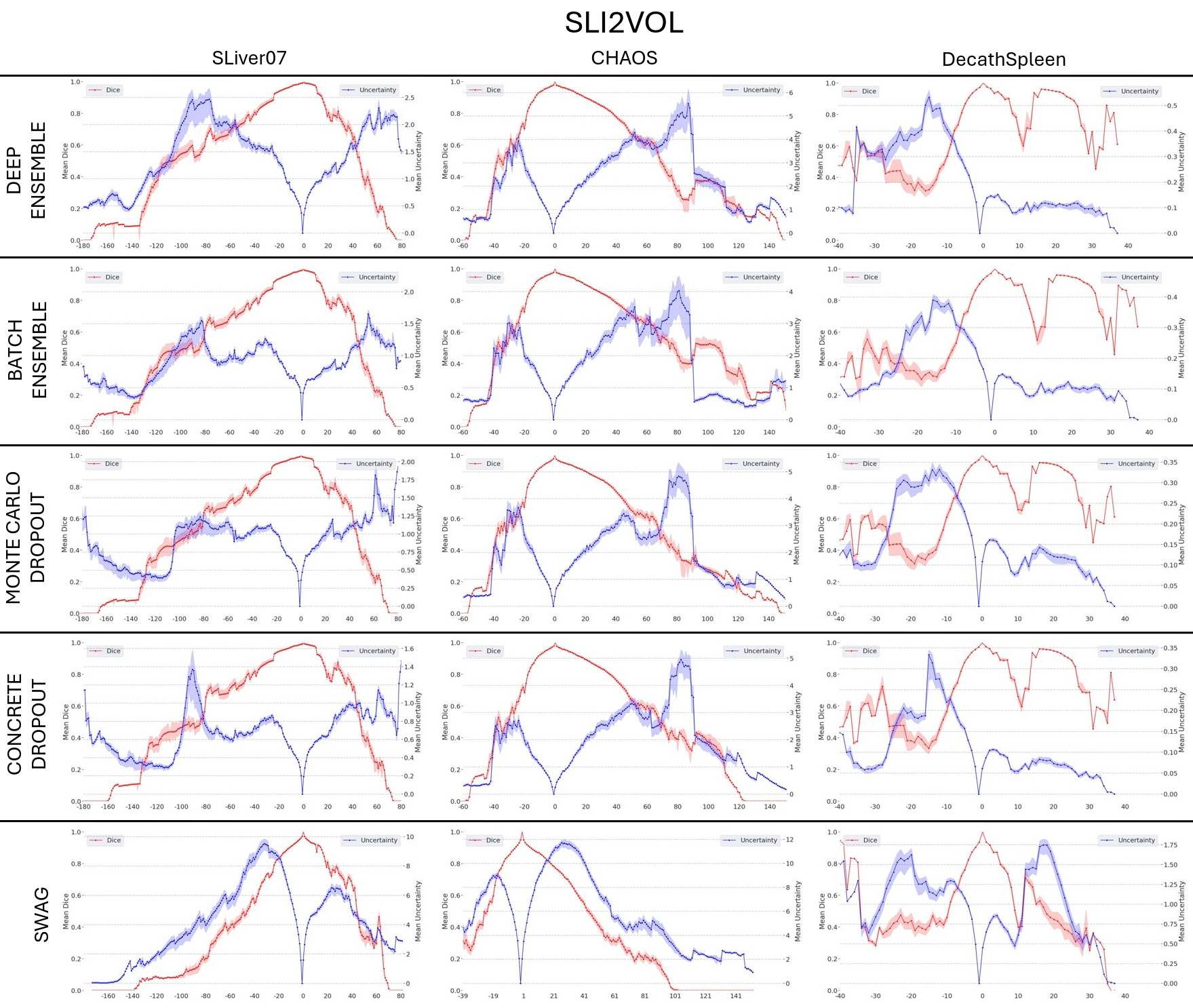}
    \includegraphics[scale=0.16]{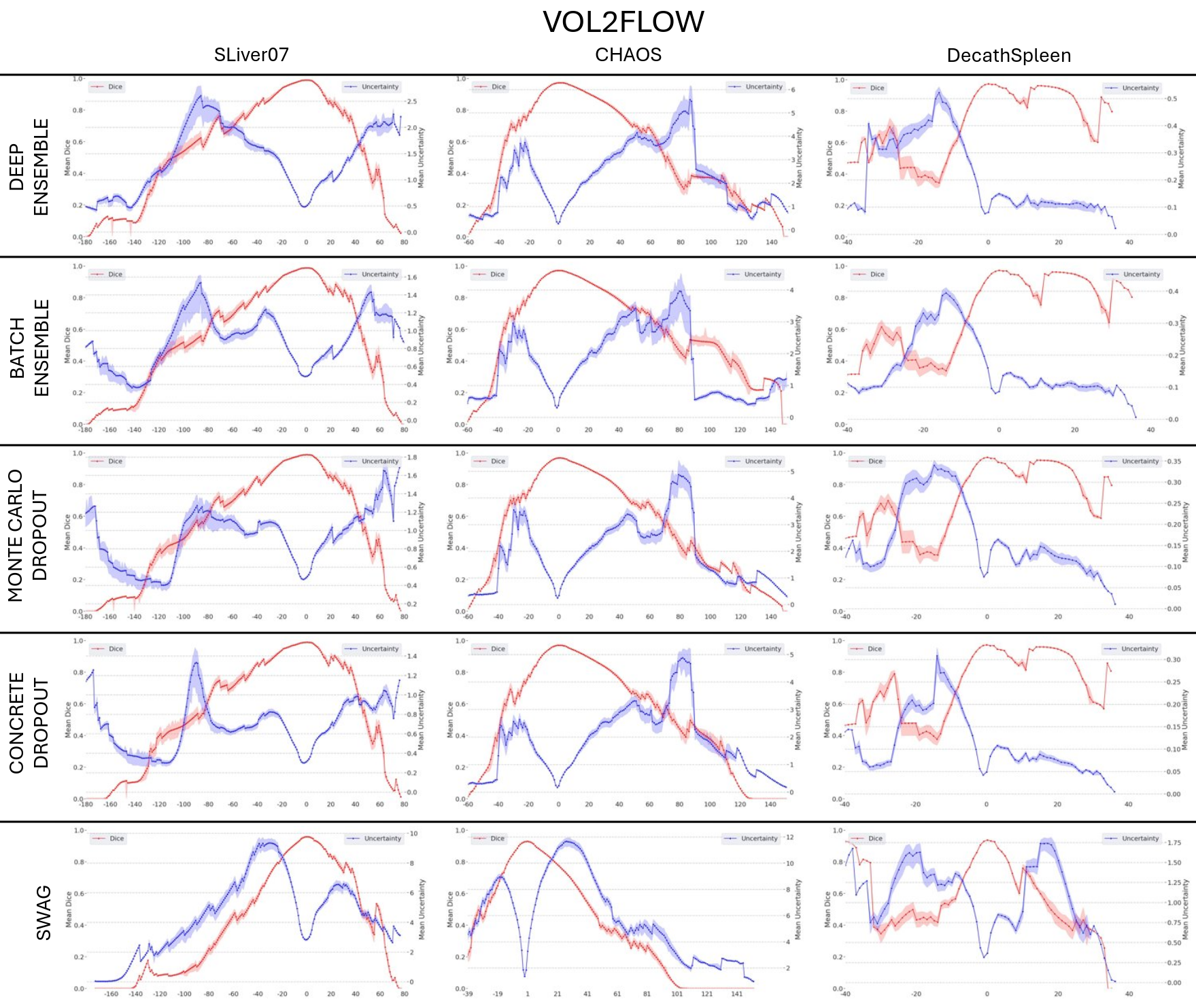}
    \caption{\textbf{Comparative analysis of variability in DSC and uncertainty metrics relative to the distance from the annotated slice} This figure presents a comparison of DSC variability and uncertainty metrics across each slice propagation method, dataset, and UQ method. A consistent trend is observed across all categories. Supplementary GIFs are provided to visually demonstrate the progression of predicted segmentations and associated uncertainties through a volume.}
    \label{fig:s2vv2f}
    \vspace{-1em}
\end{figure}

%% file: Paper-3515.bbl
\begin{thebibliography}{10}
\providecommand{\url}[1]{\texttt{#1}}
\providecommand{\urlprefix}{URL }
\providecommand{\doi}[1]{https://doi.org/#1}

\bibitem{adams2020uncertain}
Adams, J., Bhalodia, R., Elhabian, S.: Uncertain-deepssm: From images to probabilistic shape models. In: Shape in Medical Imaging: International Workshop, ShapeMI 2020, Held in Conjunction with MICCAI 2020, Lima, Peru, October 4, 2020, Proceedings. pp. 57--72. Springer (2020)

\bibitem{adams2023benchmarking}
Adams, J., Elhabian, S.Y.: Benchmarking scalable epistemic uncertainty quantification in organ segmentation. In: International Workshop on Uncertainty for Safe Utilization of Machine Learning in Medical Imaging. pp. 53--63. Springer (2023)

\bibitem{alnazer2021recent}
Alnazer, I., Bourdon, P., Urruty, T., Falou, O., Khalil, M., Shahin, A., Fernandez-Maloigne, C.: Recent advances in medical image processing for the evaluation of chronic kidney disease. Medical Image Analysis  \textbf{69},  101960 (2021)

\bibitem{arazo2020pseudo}
Arazo, E., Ortego, D., Albert, P., O’Connor, N.E., McGuinness, K.: Pseudo-labeling and confirmation bias in deep semi-supervised learning. In: 2020 International Joint Conference on Neural Networks (IJCNN). pp.~1--8. IEEE (2020)

\bibitem{bitarafan2022vol2flow}
Bitarafan, A., Azampour, M.F., Bakhtari, K., Soleymani~Baghshah, M., Keicher, M., Navab, N.: Vol2flow: Segment 3d volumes using a sequence of registration flows. In: International Conference on Medical Image Computing and Computer-Assisted Intervention. pp. 609--618. Springer (2022)

\bibitem{cai2023orthogonal}
Cai, H., Li, S., Qi, L., Yu, Q., Shi, Y., Gao, Y.: Orthogonal annotation benefits barely-supervised medical image segmentation. In: Proceedings of the IEEE/CVF Conference on Computer Vision and Pattern Recognition. pp. 3302--3311 (2023)

\bibitem{chen2017rethinking}
Chen, L.C., Papandreou, G., Schroff, F., Adam, H.: Rethinking atrous convolution for semantic image segmentation. arXiv preprint arXiv:1706.05587  (2017)

\bibitem{gal2016dropout}
Gal, Y., Ghahramani, Z.: Dropout as a bayesian approximation: Representing model uncertainty in deep learning. In: international conference on machine learning. pp. 1050--1059. PMLR (2016)

\bibitem{gal2017concrete}
Gal, Y., Hron, J., Kendall, A.: Concrete dropout. Advances in neural information processing systems  \textbf{30} (2017)

\bibitem{heimann2009comparison}
Heimann, T., Van~Ginneken, B., Styner, M.A., Arzhaeva, Y., Aurich, V., Bauer, C., Beck, A., Becker, C., Beichel, R., Bekes, G., et~al.: Comparison and evaluation of methods for liver segmentation from ct datasets. IEEE transactions on medical imaging  \textbf{28}(8),  1251--1265 (2009)

\bibitem{heller2023kits21}
Heller, N., Isensee, F., Trofimova, D., Tejpaul, R., Zhao, Z., Chen, H., Wang, L., Golts, A., Khapun, D., Shats, D., et~al.: The kits21 challenge: Automatic segmentation of kidneys, renal tumors, and renal cysts in corticomedullary-phase ct. arXiv preprint arXiv:2307.01984  (2023)

\bibitem{icsin2016review}
I{\c{s}}{\i}n, A., Direko{\u{g}}lu, C., {\c{S}}ah, M.: Review of mri-based brain tumor image segmentation using deep learning methods. Procedia Computer Science  \textbf{102},  317--324 (2016)

\bibitem{izmailov2018averaging}
Izmailov, P., Podoprikhin, D., Garipov, T., Vetrov, D., Wilson, A.G.: Averaging weights leads to wider optima and better generalization. arXiv preprint arXiv:1803.05407  (2018)

\bibitem{jungo2019assessing}
Jungo, A., Reyes, M.: Assessing reliability and challenges of uncertainty estimations for medical image segmentation. In: Medical Image Computing and Computer Assisted Intervention--MICCAI 2019: 22nd International Conference, Shenzhen, China, October 13--17, 2019, Proceedings, Part II 22. pp. 48--56. Springer (2019)

\bibitem{kataria2023automating}
Kataria, T., Rajamani, S., Ayubi, A.B., Bronner, M., Jedrzkiewicz, J., Knudsen, B.S., Elhabian, S.Y.: Automating ground truth annotations for gland segmentation through immunohistochemistry. Modern Pathology  \textbf{36}(12),  100331 (2023)

\bibitem{kendall2017uncertainties}
Kendall, A., Gal, Y.: What uncertainties do we need in bayesian deep learning for computer vision? Advances in neural information processing systems  \textbf{30} (2017)

\bibitem{kuisma2020validation}
Kuisma, A., Ranta, I., Keyril{\"a}inen, J., Suilamo, S., Wright, P., Pesola, M., Warner, L., L{\"o}yttyniemi, E., Minn, H.: Validation of automated magnetic resonance image segmentation for radiation therapy planning in prostate cancer. Physics and imaging in radiation oncology  \textbf{13},  14--20 (2020)

\bibitem{lakshminarayanan2017simple}
Lakshminarayanan, B., Pritzel, A., Blundell, C.: Simple and scalable predictive uncertainty estimation using deep ensembles. Advances in neural information processing systems  \textbf{30} (2017)

\bibitem{li2023well}
Li, W., Yuille, A., Zhou, Z.: How well do supervised models transfer to 3d image segmentation? In: The Twelfth International Conference on Learning Representations (2023)

\bibitem{maddox2019simple}
Maddox, W.J., Izmailov, P., Garipov, T., Vetrov, D.P., Wilson, A.G.: A simple baseline for bayesian uncertainty in deep learning. Advances in neural information processing systems  \textbf{32} (2019)

\bibitem{niyas2022medical}
Niyas, S., Pawan, S., Kumar, M.A., Rajan, J.: Medical image segmentation with 3d convolutional neural networks: A survey. Neurocomputing  \textbf{493},  397--413 (2022)

\bibitem{prassni2010uncertainty}
Prassni, J.S., Ropinski, T., Hinrichs, K.: Uncertainty-aware guided volume segmentation. IEEE transactions on visualization and computer graphics  \textbf{16}(6),  1358--1365 (2010)

\bibitem{ct-pancreas}
Roth, H., Farag, A., Turkbey, E.B., Lu, L., Liu, J., Summers, R.M.: Data from pancreas-ct (version 2) [data set]. The Cancer Imaging Archive. https://doi.org/10.7937/K9/TCIA.2016.tNB1kqBU  (2016)

\bibitem{ct-lymphnodes}
Roth, H., Lu, L., Seff, A., Cherry, K.M., Hoffman, J., Wang, S., Liu, J., Turkbey, E., Summers, R.M.: A new 2.5 d representation for lymph node detection in ct [data set]. The Cancer Imaging Archive. https://doi.org/10.7937/K9/TCIA.2015.AQIIDCNM  (2015)

\bibitem{shi2021inconsistency}
Shi, Y., Zhang, J., Ling, T., Lu, J., Zheng, Y., Yu, Q., Qi, L., Gao, Y.: Inconsistency-aware uncertainty estimation for semi-supervised medical image segmentation. IEEE transactions on medical imaging  \textbf{41}(3),  608--620 (2021)

\bibitem{simpson2019large}
Simpson, A.L., Antonelli, M., Bakas, S., Bilello, M., Farahani, K., van Ginneken, B., Kopp-Schneider, A., Landman, B.A., Litjens, G., Menze, B., Ronneberger, O., Summers, R.M., Bilic, P., Christ, P.F., Do, R.K.G., Gollub, M., Golia-Pernicka, J., Heckers, S.H., Jarnagin, W.R., McHugo, M.K., Napel, S., Vorontsov, E., Maier-Hein, L., Cardoso, M.J.: A large annotated medical image dataset for the development and evaluation of segmentation algorithms (2019)

\bibitem{swayamdipta2020dataset}
Swayamdipta, S., Schwartz, R., Lourie, N., Wang, Y., Hajishirzi, H., Smith, N.A., Choi, Y.: Dataset cartography: Mapping and diagnosing datasets with training dynamics. arXiv preprint arXiv:2009.10795  (2020)

\bibitem{valindria2018multi}
Valindria, V.V., Pawlowski, N., Rajchl, M., Lavdas, I., Aboagye, E.O., Rockall, A.G., Rueckert, D., Glocker, B.: Multi-modal learning from unpaired images: Application to multi-organ segmentation in ct and mri. In: 2018 IEEE winter conference on applications of computer vision (WACV). pp. 547--556. IEEE (2018)

\bibitem{wen2020batchensemble}
Wen, Y., Tran, D., Ba, J.: Batchensemble: an alternative approach to efficient ensemble and lifelong learning. arXiv preprint arXiv:2002.06715  (2020)

\bibitem{yeung2021sli2vol}
Yeung, P.H., Namburete, A.I., Xie, W.: Sli2vol: Annotate a 3d volume from a single slice with self-supervised learning. In: Medical Image Computing and Computer Assisted Intervention--MICCAI 2021: 24th International Conference, Strasbourg, France, September 27--October 1, 2021, Proceedings, Part II 24. pp. 69--79. Springer (2021)

\end{thebibliography}
